# Creep properties and deformation mechanisms of single-crystalline γ′-strengthened superalloys in dependence of the Co/Ni ratio

N. Volz[a], C.H. Zenk[a], N. Karpstein[b], M. Lenz[b], E. Spiecker[b], M. Göken[a], S. Neumeier[a]

[a] Friedrich-Alexander-Universität Erlangen-Nürnberg, Department of Materials Science & Engineering, Institute I: General Materials Properties, 91058 Erlangen, Germany

[b] Friedrich-Alexander-Universität Erlangen-Nürnberg, Department of Materials Science & Engineering, Institute of Micro- and Nanostructure Research, and Center for Nanoanalysis and Electron Microscopy (CENEM), Friedrich Alexander University of Erlangen-Nuremberg, IZNF 91058 Erlangen, Germany

______________________________________________________________________

**Abstract**

Co-base superalloys are considered as promising high temperature materials besides the well-established Ni-base superalloys. However, Ni appears to be an indispensable alloying element also in Co-base superalloys. To address the influence of the base elements on the deformation behavior, high-temperature compressive creep experiments were performed on a single crystal alloy series that was designed to exhibit a varying Co/Ni ratio and a constant Al, W and Cr content. Creep tests were performed at 900 °C and 250 MPa and the resulting microstructures and defect configurations were characterized via electron microscopy. The minimum creep rates differ by more than one order of magnitude with changing Co/Ni ratio. An intermediate CoNi-base alloy exhibits the overall highest creep strength. Several strengthening contributions like solid solution strengthening of the γ phase, effective diffusion coefficients or stacking fault energies were quantified. Precipitate shearing mechanisms differ significantly when the base element content is varied. While the Ni-rich superalloys exhibit SISF and SESF shearing, the Co-rich alloys develop extended APBs when the γ′ phase is cut. This is mainly attributed to a difference in planar fault energies, caused by a changing segregation behavior. As result, it is assumed that the shearing resistivity and the occurring deformation mechanisms in the γ′ phase are crucial for the creep properties of the investigated alloy series.

Keywords: creep; rafting; electron microscopy; superalloy; deformation mechanisms; Co/Ni ratio, solid solution strengthening, stacking fault energy

______________________________________________________________________________

# 1. Introduction

Plenty of studies on creep deformation mechanisms have been carried out on various Co- and CoNi-based alloys under different test conditions to analyze their potential compared to conventional Ni-base superalloys [1–12]. Depending on alloy composition, loading direction, temperature and applied stress, the deformation mechanisms differ significantly. For example, Suzuki and Pollock [1] reported that tensile creep deformation in the temperature range from 600 °C to 900 °C and at stresses between 240 MPa and 530 MPa is based on matrix deformation by ⟨110⟩ slip and γ′ deformation on ⟨110⟩{111} and ⟨112⟩{111} slip systems. Titus et al. [2] showed that the γ′ precipitates in a Co-base superalloy, tested at 900 °C, are sheared by a/3⟨112⟩ partial dislocations, which create superlattice intrinsic (SISF) and extrinsic (SESF) stacking faults, whereas in a CoNi-based alloy, γ′ is sheared by a/2⟨110⟩ dislocations, leaving behind anti-phase boundaries (APBs). Additionally, Eggeler et al. [3] found a fault configuration during tensile creep at 900 °C and 310 MPa where an SISF is embedded in an APB (ASA-configuration). These configurations differ significantly from the ones observed in Ni-base superalloys, which are known to deform mainly by matrix dislocation movement in the high temperature creep regime [13–16]. Of course, it has to be considered that the so-called high temperature creep regime is defined to start at lower temperatures for Co-base superalloys (e.g. 900 °C) than for Ni-base superalloys (e.g. 1000 °C). However, deformation by γ′ shearing under the formation of superlattice stacking faults was found for Ni-base superalloys as well, although at lower temperatures like 750 °C and high stresses [17]. At intermediate temperatures of 850 °C it was found that the deformation of Ni-base superalloys is dominated by matrix deformation in the early stages of creep and the resistance of γ′ against shearing is the strength-limiting factor [18,19]. When γ′ cutting occurred at these temperatures, APB coupled a/2⟨110⟩ dislocation pairs were observed [18,19].

Another important difference between Co- and Ni-base superalloys, leading to different creep properties, is the sign of the γ/γ′ lattice misfit. While most cast Ni-base superalloys exhibit a negative misfit [20–23], it is typically positive for Co-base superalloys [1,6,24,25]. The lattice misfit between the γ and γ′ phase – no matter if positive or negative – leads to a directional coarsening (rafting) of the γ′ phase during creep at high temperatures. However, the preferred direction of rafting changes with the sign of the lattice misfit. The precipitates align parallel to the external stress axis in alloys with negative misfit in compression and perpendicular to it under tensile loading. It was found for Co- and Ni-base superalloys that the rafting behavior can have beneficial effects for the creep properties, depending on the test parameters [6,7,26,27].

It was already found that the Ni content in Co-Al-W-based superalloys significantly influences the partitioning behavior of other alloying elements. The changing compositions of the γ and γ′ phases result in a different amount of solid solution strengthening, γ/γ′ lattice misfits [28–30] and planar defect energies [2,11,31,32]. As a result, the overall creep properties vary significantly between Co- and Ni-base superalloys. To address the influence of the base elements Co and Ni systematically, a model alloy series has been developed by Zenk et al. [29,30]. They investigated polycrystalline (PX) specimens of alloys with varying Co/Ni ratios. It was found that the alloying elements distribute more evenly, the lattice misfit switches from negative to positive values and the creep properties deteriorate with increasing Co content [29,30]. The aim of the present study is to analyze how the variation in microstructure and thermophysical properties, which is induced by a change of the Co/Ni ratio, influences the creep properties and deformation behavior of single-crystalline (SX) sample material. Therefore, creep tests at 900 °C and 250 MPa were performed and the resulting defect structures were characterized by scanning (SEM) and transmission electron microscopy (TEM).

## 2. Experimental procedures

### 2.1. Materials and processing

The nominal compositions of the investigated alloys are given in Table 1. Polycrystalline samples of these alloys were investigated in earlier studies in terms of thermophysical and mechanical properties and the interested reader is referred to [29,30]. It is worth noting that the alloys in those studies additionally contained boron to overcome grain boundary embrittlement. The number X in the alloy denotations represents the Co-fraction with respect to the overall content of the base elements Co and Ni, i.e. in NCX, $X = c(Co) / (c(Co)+c(Ni)) * 100$. Therefore, NC0 is a pure Ni-base alloy and NC100 a pure Co-base superalloy, whereas NC25, NC50 and NC75 are intermediate alloys with increasing Co-content.

**Table 1: Nominal composition of NC0, 25, 50, 75 and 100 in at.%.**

| NCX   | Co    | Ni    | Al   | W    | Cr   |
|-------|-------|-------|------|------|------|
| NC0   | -     | 75.00 | 9.00 | 8.00 | 8.00 |
| NC25  | 18.75 | 56.25 | 9.00 | 8.00 | 8.00 |
| NC50  | 37.50 | 37.50 | 9.00 | 8.00 | 8.00 |
| NC75  | 56.25 | 18.75 | 9.00 | 8.00 | 8.00 |
| NC100 | 75.00 | -     | 9.00 | 8.00 | 8.00 |

To study the deformation mechanisms in more detail, single-crystalline rods of these compositions were produced using the Bridgman process at withdrawal rates of 3 mm/min. EBSD measurements were used to determine the misorientation of the cast material. Based on these measurements, ⟨001⟩-oriented segments were extracted from the rods with a deviation less than 5°. Creep specimens and samples for microstructure analysis were prepared from these segments after heat-treatments. All alloys were solution annealed at 1250 °C for 24 h and aged at 900 °C for 100 h in vacuum to provide a homogeneous two-phase γ/γ′ microstructure. The cooling rate between the two heat-treatment temperatures and from 900 °C to room temperature was approximately 300 °C/h. For SEM microstructure analysis, the samples were ground to 4000 grit and mechanically polished using diamond suspensions, followed by a chemo-mechanical fine polishing (*Struers, OPS*). The microstructure was investigated using a *Zeiss Crossbeam 1450 EsB* and backscattered electron imaging (BSD). The γ′ area fraction was measured by ImageJ. From that the γ′ volume fraction was calculated according to the shape factor of the precipitates, as described in [33].

Cylindrical samples with a diameter of about 5 mm and a height of about 7.5 mm were used to perform compression creep tests at 900 °C and 250 MPa. TEM specimens were produced by precision-cutting of disks of about 200 μm thickness, followed by grinding to 2500 grit. The final thinning was done using a twin-jet polishing machine with a 60 % perchloric acid in methanol and 2-butoxyethanol electrolyte (*Struers, A3 electrolyte*). A *Philips CM200* at 200 kV and a *FEI Titan Themis³* at 300 kV were used to analyze the deformation structures in TEM bright-field (BF) and dark-field (DF) imaging and scanning TEM (STEM) mode using a high angle annular dark-field (HAADF) detector.

### 2.2. Quantification of strengthening contributions

Thermo-Calc (TC) was used in an attempt to calculate the stacking fault energies $\gamma_{SFE}$ of the matrix phase of NC-alloys. The γ compositions were experimentally determined (APT) and are given in Table 3. The model originally developed by Olson and Cohen [34], which is primarily based on the Gibbs energy difference of the fcc-γ and hcp-ε phases, was used for that:

$$\gamma_{SFE} = 2\rho_\gamma \left( \Delta G_m^{\gamma \rightarrow \varepsilon} + E_m^{str} \right) + 2\sigma_{\gamma/\varepsilon} \qquad \text{Equation 1}$$

$\gamma_{SFE}$ is the stacking fault energy, $\rho_\gamma$ the molar surface density, $\Delta G_m^{\gamma \rightarrow \varepsilon}$ the molar Gibbs energy difference between the hcp-ε and the fcc-γ phases of the same composition, $E_m^{str}$ a molar strain energy term associated with the lattice distortion around the partial dislocations and stacking fault and $\sigma_{\gamma/\varepsilon}$ is the interfacial energy between γ and ε on the {111} stacking fault habit plane. The molar surface density ρ was calculated from the molar volume of the fcc phase at 900 °C using the TCNI10 database. As the hcp phase in TCNI10 is not sufficiently well described for this purpose, $\Delta G_m^{\gamma \to \varepsilon}$ was calculated using the TTNI8 database.

The molar strain energy term $E_m^{str}$ was also calculated according to [34] from the molar volumes of the individual phases (TCNI10), the strain $\varepsilon_{33}$ along the ε-phase's c-axis associated with the γ→ε phase transformation, the Poisson ratio ν and the shear modulus μ:

$$E_m^{str} \approx \frac{2(1-\nu)}{9(1+\nu)} \mu V_m^\gamma \left( \frac{V_m^\varepsilon - V_m^\gamma}{V_m^\gamma} \right) + \frac{7-5\nu}{15(1-\nu)} V_m^{fcc} \frac{2}{3} \mu \varepsilon_{33}^2 \qquad \textbf{Equation 2}$$

As there is no data on this specific alloy system available, we estimated $\varepsilon_{33}$ to be -0.67 % based on various works investigating the martensitic transformation of Co and Co-alloys [35–40]. The values for Co range from -0.3 % to -0.8 % in these studies, however, varying $\varepsilon_{33}$ in this range does not significantly alter the findings of this study. The value ν was assumed to be 0.33. Since the shear modulus μ of the matrix composition is not known, the value for Haynes188 [41] as given by the official data sheet (61 GPa) at 900 °C was used for all alloys, since the composition is comparable to the NCX alloys and the elastic stiffness is not expected to vary much throughout the system. The strain energy determined in this way was found to be about two orders of magnitude smaller than the Gibbs energy. All values used for the variables in the calculation of the matrix stacking fault energy as well as intermediate and final results are summarized in Table 2.

Table 2: Calculated and literature values for estimating the stacking fault energies in single phase alloys of the experimentally determined matrix compositions (see Table 3) of alloys NCX aged at 900°C. Molar volumes $V_{m,\gamma}$ and $V_{m,\varepsilon}$ of the respective phases, molar surface density of the matrix phase $\rho_\gamma$, molar Gibbs energy $G_m^\varepsilon$ and $G_m^\gamma$ and their difference $\Delta G_m^{\gamma \to \varepsilon}$, Shear modulus $\mu$ of Haynes 188, Poisson ratio $\nu$, strain $\varepsilon_{33}$ along the new hcp c axis during the $\gamma \to \varepsilon$ transformation, molar strain energy $E_m^{str}$ associated with a stacking fault and the resulting stacking fault energy $\gamma_{SFE}$. All values except for the interfacial energy $\sigma_{\gamma/\varepsilon}(0K)$ correspond to a temperature of 900 °C.

| | comment | NC0 | NC25 | NC50 | NC75 | NC100 |
|---|---|---|---|---|---|---|
| $V_{m,\gamma}$ / cm³/mole | TCNI10 | 7.1585 | 7.2031 | 7.230 | 7.2809 | 7.4209 |
| $V_{m,\varepsilon}$ / cm³/mole | TCNI10 | 7.3804 | 7.3699 | 7.3650 | 7.4062 | 7.4415 |
| $\rho_\gamma$ / mole/m² | - | 2.92×10⁵ | 2.91×10⁵ | 2.90×10⁵ | 2.89×E×10⁵ | 2.85×10⁵ |
| $G_m^\varepsilon$ / J/mole | TTNI 8 | -67317 | -70702 | -71338 | -71599 | -67932 |
| $G_m^\gamma$ / J/mole | TTNI 8 | -64734 | -68174 | -69430 | -70080 | -66698 |
| $\Delta G_m^{\gamma \to \varepsilon}$ / J/mole | calc. | 2583 | 2528 | 1908 | 1519 | 1234 |
| $\mu$ / GPa | [41] | | | 61 | | |
| $\nu$ | - | | | 0.33 | | |
| $\varepsilon_{33}$ / % | [35–40] | | | -0.67 | | |
| $E_m^{str}$ | calc. | 54.02 | 33.48 | 24.22 | 21.90 | 7.69 |
| $2\sigma_{\gamma/\varepsilon}$ / mJ/m² | [42] | | | -3.4 | | |
| $2\rho\Delta G_m^{\gamma \to \varepsilon}$ / mJ/m² | - | 151 | 147 | 111 | 88 | 70 |
| $2\rho E_m^{str}$ / mJ/m² | - | 3.16 | 1.95 | 1.41 | 1.27 | 0.44 |
| $\gamma_{SFE}$ / mJ/m² | - | 151 | 146 | 109 | 86 | 67 |

The contribution of the solid solution hardening $\sigma_{ss}$ of the γ phase was estimated experimentally using the Labusch theory [43]. This theory was then modified by Gypen and Deruyttere for various alloying elements in multicomponent alloy systems [44,45]. According to this approach and the addition of Varvenne et al. [46] and Galindo-Nava et al. [47] for two-phase alloys, the strengthening contribution of the matrix phase by solid solution hardening can be calculated as:

$$\sigma_{ss} = (1 - f_\gamma) \left( \sum_i \beta_i^{3/2} x_i \right)^{2/3} \qquad \text{Equation 3}$$

The value $f_{\gamma'}$ gives the $\gamma'$ volume fraction and $(1-f_{\gamma'})$ limits the calculation of the solid solution strengthening to the $\gamma$ phase in the two phase system, since most of the dislocation activity is located in this phase, as shown later. $x_i$ is the atomic fraction of the element i in the $\gamma$ phase of the alloy and taken from APT measurements on PX sample material of the NCX alloy series derived from [30]. Later in the manuscript, the strengthening contribution assuming a single phase fcc alloy with the composition of the $\gamma$ phase and also the strengthening contribution with respect to the two phase microstructure by taking the different $\gamma'$ volume fractions into account will be shown. The constants $\beta_i$ of alloying elements $i$ were calculated according to Fleischer [48] and describe the lattice and shear modulus misfit between the solute and Ni. It can be calculated according to equation 4.

$$\beta_i = \frac{3}{2}\mu(\eta'_i + 16\delta_i)^{3/2} \qquad \textbf{Equation 4}$$

As before, the shear modulus $\mu$ of Haynes188 at 900 °C was used for all alloys. The constant $\eta'_i$ describes the difference in shear moduli and can be calculated as $\eta'_i = \frac{\eta_i}{1 + 0.5\,|\eta_i|}$ where $\eta_i = \frac{\mu_i - \mu_{Ni}}{\mu_{Ni}}$ with $\mu_i$ being the shear modulus of solute $i$ and $\mu_{Ni}$ the shear modulus of Nickel. The constant $\delta_i$ describes the difference in atomic radii and can be derived from $\delta_i = \frac{r_i - r_{Ni}}{r_{Ni}}$ where $r_i$ is the atomic radius of solute $i$ and the atomic radius of Nickel. Since the atomic radius and shear modulus of Ni and Co do not differ significantly, Ni was used as reference element for all alloys investigated. Due to this, Co was also not considered as a solid solution strengthening element, since its effect can be neglected in the reference system Ni, according to the applied model. We know that the models described above are considered to be valid only for small solute additions and the composition especially of the base element is changing significantly in our alloy series. Additionally, the solid solution strengthening effect by changing the stacking fault energy is not covered by these models. However, since the effect of Co is negligible due to the small differences regarding atomic size and shear modulus to Ni, application of the models is assumed to be reasonable for our alloys. To calculate the solid solution strengthening at the creep test temperature of 900 °C, also temperature dependent shear moduli and atomic radii were used. Shear moduli were linearly extrapolated to 900 °C from temperature-dependent measurements taken from refs. [49–53] if they were not available directly. The atomic radii of the solutes at 900 °C were taken from thermodynamic calculations using Thermo-Calc with the SGTE unary database version 5.1. All values which were used for the calculations are listed in Table 3.

**Table 3: Shear moduli $\mu_i$ and atomic radii $r_i$ at 900 °C used for the calculation of the solid solution strengthening and the experimentally determined (APT) composition of the γ phases $x_i$, taken from [30]. Shear moduli of solutes were linearly extrapolated from refs. [49–53] if not available at 900 °C. Atomic radii were calculated using Thermo-Calc (TCNI10 database). The shear modulus of Haynes188 [41] was used, since it was not available for the NCX matrix composition. The calculated values for $\beta_i$ are also presented.**

|  | Co | Ni | Al | W | Cr | Haynes188 [41] |
|---|---|---|---|---|---|---|
| $\mu_i$ (900°C) / GPa | 49.2 | 56.4 | 12.2 | 143.3 | 101.0 | 61 |
| $r_i$ (900°C) / nm | 0.126 | 0.126 | 0.147 | 0.144 | 0.131 | - |
| $x_i$ (NC0) / at.% | - | 76.0 | 3.9 | 8.1 | 12.0 | - |
| $x_i$ (NC25) / at.% | 25.1 | 49.9 | 3.8 | 7.8 | 13.4 | - |
| $x_i$ (NC50) / at.% | 45.6 | 30.5 | 4.9 | 6.8 | 12.2 | - |
| $x_i$ (NC75) / at.% | 59.7 | 16.1 | 7.2 | 6.1 | 10.9 | - |
| $x_i$ (NC100) / at.% | 75.8 | - | 8.7 | 6.1 | 9.4 | - |
| $\beta_i$ / MPa/at.%$^{2/3}$ | 5.2 | - | 524.5 | 505.8 | 125.0 | - |

To quantify diffusional effects during creep, a model derived by Zhu et al. [54] was used to calculate the so called effective diffusion coefficient $D^{eff}$. This empirical parameter can be described as the average mobility of vacancies and is defined by:

$$D^{eff} = D_0^{eff} exp\left(\frac{-Q_{eff}}{RT}\right) \qquad \textbf{Equation 5}$$

where R is the universal gas constant and $D_0^{eff}$ and $Q_{eff}$ are the frequency factor and the activation energy of the diffusing alloying elements, respectively, equivalent to the diffusion of a single solute. They can be calculated as:

$$D_0^{eff} = \left(\sum_i \frac{c_i}{D_0^{i,base}}\right)^{-1} \qquad \textbf{Equation 6}$$

$$Q_{eff} = Q_{base} + \sum_i c_i Q_{i,base} \qquad \textbf{Equation 7}$$

Thus $D_0^{eff}$ is the harmonic mean of the frequency factors $D_0$ of the solutes i in the base element of an alloy. The effective activation energy $Q_{eff}$ is calculated from the activation energy for self diffusion in the base element $Q_{base}$ and the activation energies for diffusion of the solute $i$ in the base element, weighted according the elemental content $c_i$.

## 3. Results

### 3.1. Initial state

The two-phase γ/γ′ microstructures of the five alloys NC0, 25, 50, 75 and 100 after solution and aging heat-treatment are shown in Figure 1. While NC0 shows mainly cubic precipitates, the γ′ phase approaches a more globular morphology with increasing Co content, until a Ni/Co ratio of 1:1 (NC50) is reached. When the Co content is further increased, the precipitates start getting cubic again. This indicates a change of the γ/γ′ lattice misfit from likely negative in the Ni-rich alloys to nearly zero in NC50 (see Figure 5c), and to positive values in the Co-rich alloys NC75 and NC100, which confirms the findings of Zenk et al. who investigated the polycrystalline variants of these alloys [30].

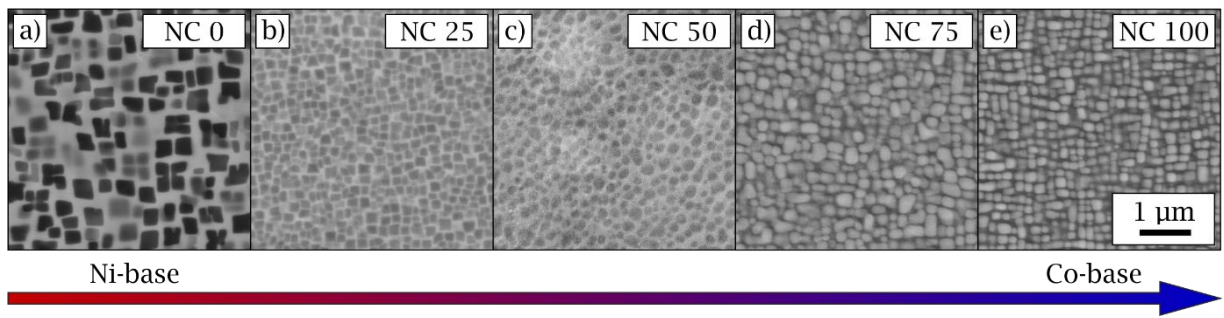

**Figure 1: SEM (BSE) images of the microstructures of the NCX alloys after solution and aging heat-treatments at 1250 °C for 24 h and at 900 °C for 100 h, respectively.**

Additionally, the contrast inversion between precipitates and matrix indicates a change in the elemental segregation between the two phases γ and γ′ (compare Figure 1 a and e). This is consistent with the findings in [30] where the heavy element W, showing strong electron-back-scattering, changes its preferred partitioning from the matrix on the Ni-rich side to the precipitate phase on the Co-rich side. All other alloying elements do not change their partitioning preference, however, they distribute more equally with increasing Co content [30].

## 3.2. Creep properties

The compression creep properties of the five investigated alloys at 900 °C and 250 MPa are shown in Figure 2a. The minimum creep rates, evaluated from these data, are shown in Figure 2b. NC25, a Ni-rich alloy, exhibits the best creep properties. A minimum creep rate of $1.5\times10^{-8}$ 1/s is reached for NC25 at about 0.5 % plastic strain, followed by a slight continuous softening. The pure Ni-base alloy NC0 shows a more constant strain rate during the creep test. The intermediate alloy NC50 exhibits a sharp minimum at about 0.2 % plastic strain and, subsequently, a significant softening, similar to NC25. The minimal strain rate, however, only reaches roughly $4.0\times10^{-8}$ 1/s, which is significantly higher compared to NC25. The deformation behavior seems to change completely for the Co-rich alloys NC75 and NC100. Both alloys exhibit a double minimum curve shape including a local minimum at small plastic strains, followed by an increase in strain rate and again a hardening to a global minimum strain rate at 6 % and 5 % plastic strain, respectively. In summary, the Ni-rich alloys exhibit significantly better creep properties at the test parameters of 900 °C and 250 MPa.

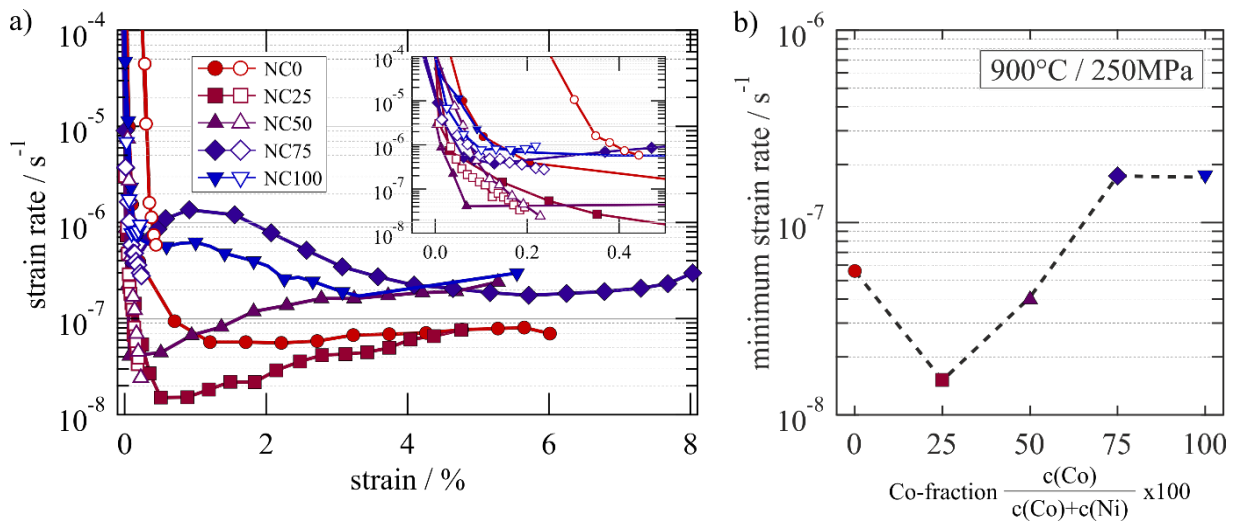

**Figure 2: a) Compressive creep properties of the NCX alloys at 900 °C and 250 MPa. Repeated tests were interrupted at strain levels of about 0.2-0.5 % plastic strain, which can be seen from the inset figure that is a magnification of the low strain part of the graph. b) Minimum strain rates evaluated from the data shown in a).**

## 3.3. Directional coarsening during creep

The microstructures after compressive creep tests to plastic strain values of 5-8 % are shown in Figure 3. NC50 does not exhibit any directional coarsening (Figure 3 c) since there is no driving force for this process due to the near-zero lattice misfit [30]. The rafting in the Ni-rich alloys NC0 and NC25 aligns parallel to the external compressive load axis (see Figure 3 a and b), which is expected for negative γ/γ′ lattice misfit alloys. In NC0, the horizontal channels have not yet closed entirely and the directional coarsening in NC25 seems to have advanced further, despite the fact that the lattice misfit in NC0 is larger and a more pronounced rafting would be expected. However, this can be explained by the test duration: while NC25 was exposed to the test conditions for about 430 h, NC0 was only tested for 240 h. Since rafting is a diffusion-controlled process, the longer creep test results in a more pronounced directional coarsening [55].

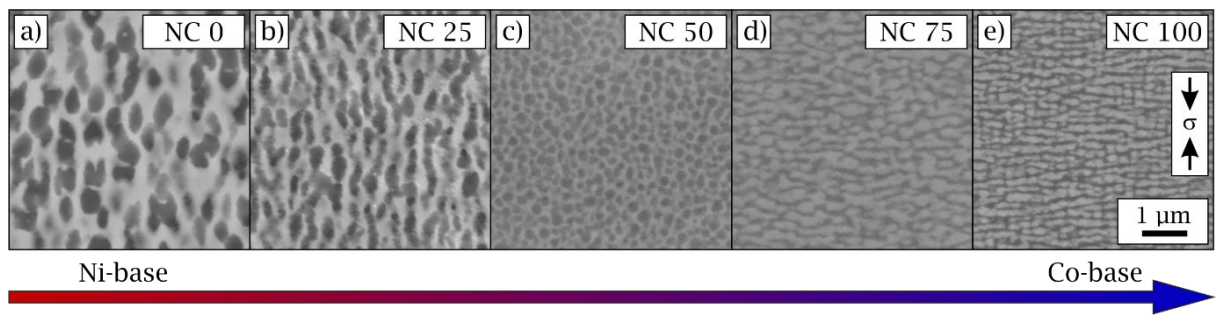

**Figure 3: SEM (BSE) images of the microstructures after creep at 900 °C and 250 MPa to 5-8% plastic strain showing the directional coarsening of alloys NC0, NC25, NC75 and NC100 and the non-directional coarsening of alloy NC50.**

The rafted γ′ microstructure in the alloys with positive lattice misfit, NC75 and NC100, is aligned perpendicular to the external stress (see Figure 3 d and e). Of these two, NC100 exhibits a considerably more evolved raft-microstructure due to the higher lattice misfit.

## 3.4. Deformation mechanisms

All creep tests at 900 °C and 250 MPa were repeated and interrupted at a plastic strain of about 0.2 % to 0.5 % (see Figure 2) to study the active deformation mechanisms in the early stages of creep. The corresponding TEM micrographs from [001] cross-sections extracted perpendicular to the stress axis are shown in Figure 4.

NC0, the Ni-base alloy, predominantly shows matrix deformation at a plastic strain of about 0.5 % (Figure 4 a,f), which is typical for Ni-base superalloys in this creep regime [18,56–58]. Most of the dislocations form networks around the γ′ precipitates as their propagation is effectively hindered by the precipitates. Similar observations of dense dislocation networks at the γ/γ′ interface have also been reported for other Ni-base superalloys [13,59–62]. Only occasionally, the γ′ phase is cut by partial dislocations, resulting in the formation of SISFs. The character of the SFs was characterized by analyzing fringe contrasts in dark-field micrographs. This mechanism also holds true for later stages of creep. The sample crept to about 6 % plastic strain shows predominantly matrix dislocations, which are surrounding the γ′ precipitates (Figure 4 k). Cutting of γ′ and the formation of superlattice stacking faults is observed only occasionally.

Shearing of the γ′ precipitates under the formation of SISFs can also be observed in NC25 after a deformation of only 0.2 %, however, deformation in the matrix via channel dislocation glide seems to be the dominant mechanism, too (Figure 4 b,g). It is worth recognizing that whenever cutting occurs in NC25, the stacking faults extend over several precipitates (but are interrupted in the matrix phase between them), which is in contrast to the mechanism observed in NC0. This indicates a slightly reduced precipitate stacking fault energy in this alloy compared to NC0, leading to a larger dissociation distance of partial dislocations. Additionally, a mechanism recently described by Eggeler et al. [3] in tensile crept specimens of a Ni-containing Co-base alloy at 900 °C could be observed: an SISF embedded in an APB (labelled as ASA configuration, Figure 4 b). When deformation proceeds to higher strains, the mechanism does not change significantly (Figure 4 l). Besides a higher amount of matrix and interfacial dislocations, also the frequency of cutting events is increasing, however, the resulting planar defects remain SISFs and the ASA-configurations.

NC50 predominantly shows matrix deformation as well, however, sometimes γ′ is sheared and stacking faults extending over several precipitates can be observed (Figure 4 c,h). In contrast to NC0 and NC25, the alloy NC50 mostly exhibits superlattice extrinsic stacking faults (SESF), when dislocations shear through a precipitate. The micrograph after plastic deformation of about 5 % does not show additional effects (Figure 4 m). The main deformation is located in the matrix phase and extended stacking faults are observed, however, with a higher density.

The Co-rich alloy NC75 shows another deformation mechanism in the early creep stage. Again, the highest dislocation activity is observed in the γ matrix, however, when dislocations shear into the precipitates, extended APBs are formed (Figure 4 d, i). These sometimes extend over several precipitates. This was also found in CoNi-based superalloys during tensile creep at 900 °C, where cutting of a/2⟨011⟩ dislocations were observed creating the APBs [10]. It is also possible that these APBs originally formed as ASA-configurations, however, the SISFs get fully transformed into APBs. The sample crept to higher strains also shows extensive cutting under the formation of APBs accompanied by matrix deformation (Figure 4 n).

In the sample with lower creep strain, alloy NC100 exhibits matrix dislocations moving in pairs with a significant splitting distance (Figure 4 e,j). When shearing of γ′ occurs, even the individual superpartials dissociate to a 4-fold splitting. One of these events can be observed in Figure 4 e. This observation corroborates the assumption of low planar fault energies. At the later stage of deformation the APBs seem to be more extended, comparable to NC75 (Figure 4 o). Stacking faults were observed occasionally at this stage. Nevertheless, the main deformation still takes place in the matrix phase.

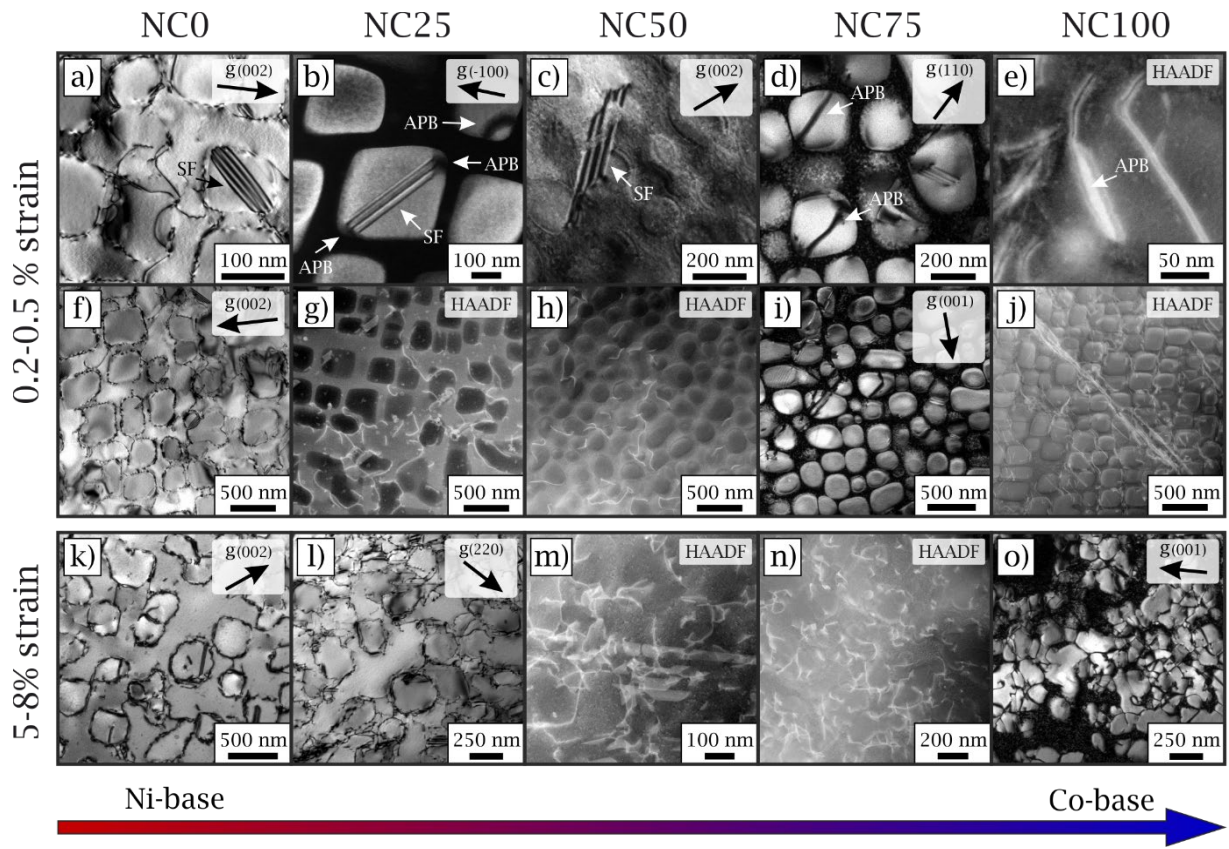

**Figure 4:** Deformation mechanisms of NC0, NC25, NC50, NC75 and NC100 observed by (S)TEM after compressive creep tests at 900 °C and 250 MPa. All micrographs are taken from ⟨001⟩ cross sections.

## 4. Discussion

The creep properties and deformation mechanisms of the NCX alloys differ significantly with the variation of the Co/Ni ratio. According to the TEM investigations, most of the deformation is located in the γ phase, which seems to allow a discussion of the creep properties in terms of solid solution strengthening and directional coarsening, however, all microstructural features like γ′ volume fraction, defect energies or diffusion properties have to be considered as well. Furthermore, interesting differences with the changing Co and Ni content also occur in early stages of creep, whenever the γ′ precipitates are sheared by dislocations.

### 4.1. Strengthening contributions

In general, the Ni-rich alloys show better creep properties compared to the Co-rich ones (see Figure 2 a and b). This might be explained by the more even distribution of alloying elements on the Co-rich side, which is known from Zenk et al. [30]. Especially W and Cr are strongly enriched in the γ matrix in NC0 and NC25, while the segregation tendency towards γ decreases with increasing Co content. This could lead to an enhanced solid solution strengthening effect in γ for these two Ni-rich alloys compared to the Co-rich ones. To prove that, the solid solution strengthening of the γ compound at 900 °C was calculated using a thermodynamic approach and the combined models of Fleischer [48], Gypen and Deruyttere [44,45] and Galindo-Nava et al. [47], as shown in Figure 5a.

The calculated strengthening contribution, weighted by the γ′ fraction in each alloy (which is also discussed below), decreases systematically with increasing Co content. Among the two Ni-rich alloys, NC25 outmatches the pure Ni-base superalloy NC0 in creep resistance. It is likely that the addition of Co in NC25 also acts as a solid-solution strengthener, since it is enriched in the γ phase. It is, however, not segregating as strongly as Cr and also the beneficial effects cannot be seen in the calculations, since this was not taken into account by the model presented in chapter 2.2. However, the calculations of solid solution strengthening fit to the results of the creep tests in the way that the creep strength decreases with increasing Co-content. Interestingly, the trend of solid solution strengthening is reversed when only the γ phase is considered and not weighted by the increasing γ′ volume fraction in the Co-rich alloys. These findings imply that the solid solution strengthening is mainly influenced by the fraction of matrix phase. While the strength of the pure γ phase slightly increases with increasing Co-content, the effect for the two-phase alloy vanishes since the γ′ fraction increases and the γ fraction decreases accordingly. However, both calculations cannot fully explain the creep properties of the investigated alloy series.

The solid solution constants $\beta_i$ we calculated using the model by Gypen and Deruyttere [44,45] and used for the estimation of the solid solution strengthening are listed in Table 3 and amount to 525, 506 and 125 MPa/at.%$^{2/3}$ for Al, W and Cr, respectively. These values indicate that Al and W are considered as nearly equal solid solution strengtheners at 900 °C and both have a higher impact compared to Cr in the Ni reference system. Recently, Wang et al. [63] also reported W to be a good solid solution strengthener in Ni binary systems while this is not the case for Cr, which is in good agreement with our findings. The high solid solution strengthening character of Al at 900 °C calculated in our study is mainly caused by the stronger temperature dependency of the atomic radius compared to Ni and the other solutes, which was derived by Thermo-Calc. Consequently, the increasing Al content in the γ phase with increasing Co-content dominates the solid solution strengthening in our calculations, since the W content is even slightly reduced.

However, the $\beta_i$ values differ significantly from the ones calculated by Galindo-Nava et al. [47] for identical solutes. While the value for Al we find is significantly higher compared to ref. [47], $\beta_i$ of W and Cr are much smaller. Two reasons may cause this effect. First, we wanted to calculate the solid solution strengthening at 900 °C and thus linearly extrapolated the shear moduli of the individual elements to high temperatures and computed theoretical atomic radii at 900 °C using Thermo-Calc. As a result, the differences with Galindo-Nava et al. [47] are reasonable, since they calculated solid solution strengthening assuming shear moduli and atomic radii of the solutes to be not at test temperature. Second, the atomic radii used for the calculations are different, especially the one of Al which was assumed to be 0.143 nm at room temperature (0.147 nm at 900 °C) compared to 0.124 nm in [47] (0.128 nm at 900 °C, assuming identical thermal behavior). A detailed discussion on the chosen atomic radii and the resulting differences can be found in part A of the supplementary material.

Additionally to the here presented calculations, the solid solution strengthening was also calculated using Thermo-Calc, where a different model is implemented. The qualitative results are similar to our findings, showing an increase in solid solution strengthening with increasing Co-content. This is described and discussed in detail in part B of the supplementary material.

Although the TEM investigations reveal pronounced deformation in the matrix phase, its solid solution strengthening estimates cannot fully explain the creep behavior and other factors must be at play. Some of the microstructural and thermophysical properties of the alloy series that could help explaining the experimental findings are shown in Figure 5 and discussed in the following.

Figure 5b illustrates the $\gamma'$ volume fraction of the NCX alloys as a function of the Co-content. The precipitate fraction is steadily increasing with increasing Co-content. This is different from the findings in [30] where a maximum of the $\gamma'$ volume fraction in polycrystalline material was reported for NC75. However, the difference between NC75 and NC100 is small as it is also the case in our study. From NC0 to NC25 the volume fraction is increasing by about 15 %, which might explain the better creep properties of this alloy [64], even if the solid solution strengthening is less pronounced according to the calculations. However, the creep properties are not improved further with increasing $\gamma'$ volume fractions, even though this would be expected for Ni-base [64] and Co-base superalloys [5]. Therefore, further properties have to be considered.

It was reported in literature, that fully developed γ′ rafts lead to a strengthening effect [6,7,26,27,55]. Although a slight directional coarsening was observed in NC0 and NC25, no direct effect can be attributed to the orientation of the rafts with respect to the external load. According to the micrographs in Figure 3a and b, the horizontal γ channels are not completely closed and presumably the forming rafts do not act as effective obstacles. Furthermore, the γ′ precipitates of the Co-base alloys reported in [7,55], where a strengthening by rafting was found during compressive creep, form plate-like morphologies during directional coarsening. For the Ni-base alloys with negative lattice misfit (NC0 and NC25), however, rod-like shapes were found in the samples crept under compression. This difference in morphology and a less pronounced rafting possibly also explains the absence of a positive effect of the rafting in NC0 and NC25. The Co-rich alloys, NC75 and NC100, exhibit a double minimum in strain rate during the creep test. This behavior is attributed to the pronounced N-type rafting, as previously described in the literature for Co-Al-W-Ta alloys [6,7]. When the vertical γ channels close and no extensive γ′ shearing occurs, the dislocations have to bypass the precipitates by glide and climb on longer paths, which leads to a measurable strengthening effect. This effect vanishes at later stages of creep when the γ′ phase coarsens and γ′ shearing becomes more pronounced.

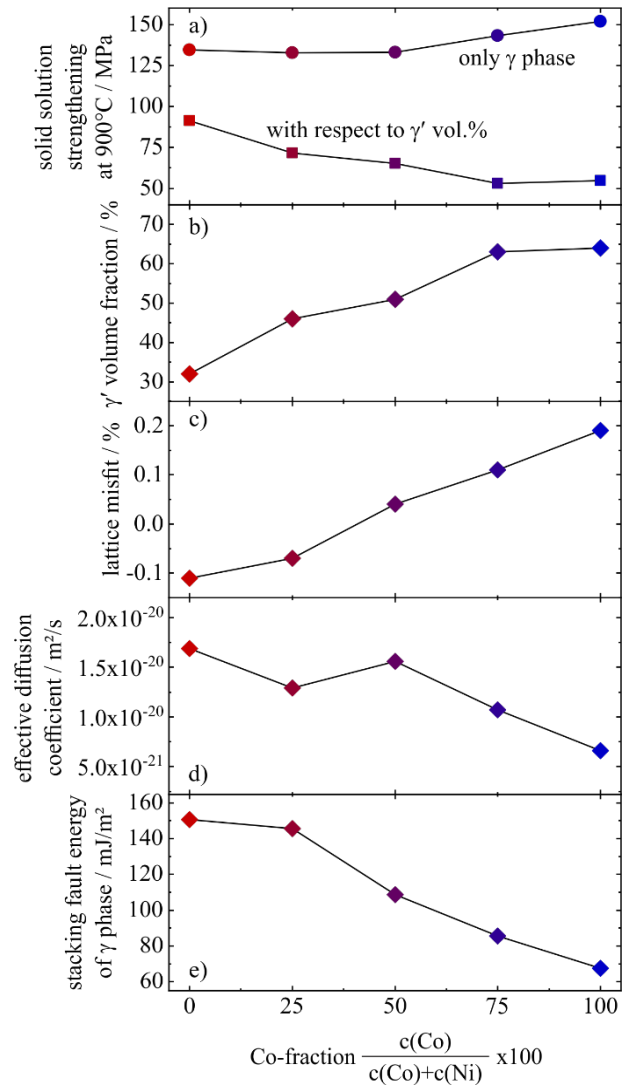

**Figure 5: Strengthening contributions of the NCX alloy series as measured or calculated. a) Solid solution strengthening of the γ phases at 900 °C calculated after [44,45,47] for the pure γ phase composition (spheres) and with respect to the γ′ volume fraction (squares), b) γ′ volume fraction evaluated from micrographs, c) γ/γ′ lattice misfit measured on polycrystalline samples in [30], d) effective diffusion coefficient of the γ phase and e) stacking fault energy of the γ phase composition as calculated using Thermo-Calc.**

The γ/γ′ lattice misfit, which also determines the morphology of the precipitates of the NCX alloys, is given in Figure 5c, as measured on polycrystalline samples by Zenk et al. [30]. It was already shown for example by Grose and Ansell [65] that higher coherency stresses can improve the mechanical properties. Assuming this, NC100 should obtain the highest strengthening contribution due to the highest lattice misfit whereas NC50, which exhibits almost globular precipitates, might then exhibit the lowest contribution. Since the Co-rich alloys show significantly lower creep strength at the tested conditions compared to the Ni-rich alloys, the strengthening by coherency stresses is also not the dominant mechanism in this alloy series. However, the lattice misfit also determines the morphology of the γ′ precipitates. The near-zero misfit of NC50 results in globular precipitates which might be unfavorable [66] and possibly explain why the creep properties of this alloy are worse compared to NC25 with a higher misfit. However, the more cubic shape of the Co-base alloys do not result in better creep properties either.

It is known that diffusion is an additional key parameter during high temperature deformation. For example the directional coarsening or the dislocation motion are strongly affected by diffusion properties [67–70]. Therefore, we calculated the effective diffusion coefficients $D^{eff}$ for the γ matrix compositions of the NCX alloys using a model derived by Zhu et al. [54] as described in the appendix. The results are shown in Figure 5d. It can be seen, that the effective diffusion coefficient is significantly lowered with increasing Co-content. The decrease from NC0 to NC25 followed by a slight increase to NC50 fits very well to the minimum strain rates observed during creep. This local minimum in the effective diffusion coefficient is most certainly caused by the enrichment of Cr in the γ phase, which has a positive effect on activation energy and frequency factor. When the Co-content is further increased, the model predicts an ongoing decrease of $D^{eff}$, although the concentration of all solutes in γ, except for Al, is decreasing. Thus, it can be stated that the diffusivity of the base elements Co and Ni is the dominating mechanism, since their high content outmatches the effect of the minor solutes. According to the model, a high Co content is beneficial, since $D_0^{Co,Co}$ is smaller and $Q^{Co,Co}$ is larger, compared to the equivalent values of Ni, independent of the assumptions made for the intermediate compositions. Additionally, the diffusivity of the solutes in Co was found to be slightly lower than in Ni [71]. However, the experimentally observed creep resistance is decreasing when the Co-content in NC25 is further increased. Therefore, also the diffusional properties of the matrix phase cannot fully explain the creep behavior, yet.

As a further strengthening contribution, we considered the stacking fault energy of the γ matrix, suggesting that a high stacking fault energy and therefore a small dissociation distance of partial dislocations promotes recombination and cross-slip. As a consequence, a high stacking fault energy $\gamma_{SFE}$ is assumed to be disadvantageous for the creep properties compared to low $\gamma_{SFE}$. The stacking fault energies of the matrix compositions (given in Table 3) of the NCX alloys were calculated using Thermo-Calc. The results are illustrated in Figure 5e. The graph shows a steady decrease of the stacking fault energy $\gamma_{SFE}$ with increasing Co-content. This would suggest easier cross-slip of dislocations on the Ni-rich side due to smaller splitting distances of the partial dislocations. Enhanced cross-slip would result in lower creep strength. However, the trend in the investigated alloy series is exactly the opposite. The creep properties are actually better in the Ni-rich alloys, although TEM investigations revealed dominant deformation in the matrix phase and TC predict the stacking fault energy to be higher.

Consequently, it is evident, that none of the strengthening contributions described above can explain the variation of creep behavior among the NCX alloys alone. Even though the matrix properties would promise better creep performance of the Co-rich alloys, the opposite trend is observed. A conclusion from this might be that the matrix properties are less important on the Co-rich side of the system: if a partial dislocation is not forced to recombine to cross-slip or climb to bypass a precipitate because the planar fault energies are low, neither SFE nor diffusivity in the matrix will be the key factor for the creep behavior. At the same time, the solid solution strengthening as the remaining contribution we considered is not increasing strongly enough to counterbalance the negative impact that the hypothesized decreased shear resistance of γ′ has.

## 4.2. Deformation mechanisms

As described in section 3.4 and shown in Figure 4, the dislocation activity in γ′ is relatively low, indicating that the main influences on the creep properties are the matrix strength and directional coarsening. However, the different properties of the matrix phase in the investigated alloys discussed above could not properly explain the creep properties and would even predict the opposite trend. We therefore assume that the planar defect energies in γ and γ′ and the associated deformation mechanisms play a key role for the overall creep properties of the alloys. Since this has not been quantified, yet, the formation of the different defect configurations will be discussed only qualitatively in the following.

In NC0, γ′ shearing and the formation of SISFs could be observed only occasionally, with increasing frequency at later creep stages. Similar behavior was observed for NC25 and is shown in Figure 4 b, g and l. Interestingly, the formation of an ASA-configuration was found, which was up to now only reported for Co- and CoNi-based alloys in tensile creep [3,4,9,72]. This configuration was now also confirmed to occur in a negative misfit Ni-based alloy creep-deformed under compression. According to Eggeler et al. [3], this configuration is formed as follows: a leading $a/3\langle 112\rangle$ super partial dislocation (formed by reaction of two matrix dislocations with dissimilar Burgers vectors) shears through γ′ and creates a SISF extending across the whole precipitate. The trailing $a/6\langle 112\rangle$ Shockley partial dislocation follows and enters the γ′ precipitate from all sides, partially transforming the SISF into an APB. As result, the trailing partial forms a loop separating the SISF (inside the loop) from the surrounding APB, both located on the {111} slip plane. Since the APB energy is lower on the {001} planes, the APB migrates from {111} to {001} [3]. It is assumed that the ASA-configuration in NC25 forms in a similar way. As marked in Figure 4 b, in some precipitates extended APBs were found in the early creep stages. It is assumed that these are former ASA-configurations where the whole SISF is transformed to an APB which subsequently migrates onto the {001} planes.

The alloy with a Co/Ni ratio of 1:1, NC50, exhibits extended stacking faults over several γ′ precipitates and intermediate channels. Both SISFs and SESFs were found in this alloy, indicating that the addition of Co affects the defect energies significantly. The formation of SESFs involves glide of two identical Shockley partial dislocations on adjacent glide planes and a successive reordering process, which is most likely also happening in NC50 [67,73–76]. However, the often described growth of SESFs into microtwins in the later creep stages was not observed in NC50.

In the Co-rich alloy NC75, APBs were observed (see Figure 4 d, i, n), indicating again a change in the planar defect energies and other dislocation reactions. Similar mechanisms were reported by Eggeler et al. [10] for Co- and CoNi-base alloys in tensile creep tests also at 900 °C. They attribute the formation of APBs to the shearing of γ′ by a/2⟨011⟩ dislocations. Shearing via APB coupled dislocation pairs was also shown for Ni-base superalloys [18,19]. However, the splitting distance is significantly smaller in those studies and does not span entire precipitates, as it is the case in NC75. It was already determined by Okamoto et al. [77] that the APB energy on the {111} planes of single-phase $Co_3(Al,W)$ is nearly 40 % lower compared to $Ni_3Al$, which fits well to our observations that APB formation is preferred in the Co-rich alloys.

The pure Co-base alloy NC100 exhibits a 4-fold dislocation dissociation when γ′ shearing occurs (see Figure 4 e, j). The middle part was also identified as an APB, indicating that the two superpartial dislocations of type a/2⟨011⟩ further dissociate into individual Shockley partial dislocations of type a/6⟨112⟩. Additionally, dislocations moving in pairs were observed in the γ channels. This could not be found in any of the other alloys.

It was already found earlier that Co and Cr additions to fcc-Ni reduce $\gamma_{SFE}$ of binary alloys [78]. This is even more pronounced with increasing solute content. Our calculations are in very good agreement with these findings since the SFE is calculated to be significantly reduced by adding more Co. Furthermore, we know that the segregation of Co and Cr to the γ phase is less pronounced in the Co-rich alloys [30]. Consequently, the content of these elements in the γ′ phase is higher compared to the Ni-rich alloys, which might then affect the planar fault energy of the precipitate phase as well. From the TEM investigations we know that shearing of γ′ is more pronounced in the Co-rich alloys, however, it can not be clarified whether this is caused by the general difference in stoichiometry of $Ni_3Al$ compared to $Co_3(Al,W)$ or by the addition of Cr or any other reason. In any case, enhanced shearing of γ′ by dislocations in the Co-rich alloys deteriorates the overall creep properties since the obstacle effect of the γ′ precipitates is diminished.

## 5. Summary and Conclusion

The properties discussed above imply that the changing creep behavior of the NCX alloys cannot be attributed directly to the changing Co/Ni ratio. Rather, it is necessary to uncover which properties are changing when the base element content is varied. It was already found that the Co/Ni ratio influences the partitioning behavior of the other alloying elements, which results in changing γ/γ′ lattice misfits [30]. In our study, we also calculated that the solid solution strengthening contribution of the γ matrix decreases with increasing Co-content, since it is dominated by the decreasing γ fraction. A quantification of the precipitation strengthening contribution to compare the NCX alloys could not be performed. Since the deformation mechanisms are not fully understood and the assumption of different parameters like fault energies could not be made, commonly used models could not be applied. Additionally, we found that the different solid solution strengthening contribution, accompanied by the inversed rafting behavior due to the opposite lattice misfit, results in a significantly altered compressive creep behavior. The deformation structures were characterized and interesting differences in the defect configurations were reported. The results imply that the variation of the γ′ planar fault energies with the changing Co/Ni ratio are the primary reason for the observed trend in creep properties. However, further work is needed to analyze the defects in more detail and to quantify their influence.

In brief summary, the creep properties and deformation behavior of a single crystalline alloy series 75(Co/Ni)-9Al-8W-8Cr, designed to map the transition from γ′-strengthened Ni-base to Co-base superalloys, was investigated at 900 °C and 250 MPa. To explain the creep properties, different strengthening contributions were quantified using existing models and Thermo-Calc. We conclude that just changing the base element influences several material properties, which creates a rather complex alloy series. Nevertheless, we could evaluate different characteristics and the following conclusion can be stated:

- The Ni-rich but Co-containing alloy NC25 exhibits the best creep properties and the creep strength significantly decreases with increasing Co-content.

- The Co-rich alloys NC75 and NC100 show a double minimum creep behavior due to a temporary strengthening by directional coarsening of the γ′ phase perpendicular to the external load.

- The partitioning behavior of all alloying elements is crucial for the mechanical properties since especially W and Al are considered as strong solid-solution strengtheners at 900 °C in this alloy series. The partitioning behavior is mainly influenced by the Co/Ni ratio.

- None of the common strengthening contributions like solid solution strengthening, γ′ volume fraction, γ/γ′ lattice misfit, diffusion coefficients or stacking fault energies alone could explain the changing creep behavior of the investigated alloy series. It is assumed that the shearing resistance of the precipitates and the deformation mechanisms play the key role in the overall creep properties.

- The deformation mechanisms change significantly with a variation of the Co/Ni ratio, especially when γ′ deformation is considered. With increasing Co-content, the γ′ cutting mechanisms change from SISF-shearing over ASA-shearing to SESF-shearing and finally APB-shearing. These changes are attributed to dramatic variations in the energies of the various types of possible planar faults in these alloys.

**Acknowledgment**

The authors acknowledge funding by the Deutsche Forschungsgemeinschaft (DFG) through projects A7 and B3 of the collaborative research center SFB/TR 103 "From Atoms to Turbine

Blades – a Scientific Approach for Developing the Next Generation of Single Crystal Superalloys".

**Appendix - Calculation of effective diffusion coefficients**

To calculate the effective diffusion coefficients of the γ compositions, a model derived by Zhu et al. [54] was used, as described in the experimental part of the manuscript. According to their model, the frequency factors $D_0$ and activation energies $Q$ for self-diffusion of the base elements and of the solutes in the base elements have to be known. This can easily be done for the alloys NC0 and NC100, which are pure Ni-base or Co-base alloys, respectively, since these parameters were already determined by other groups. However, the scope of our manuscript was to present changes induced by a change of the base element content and therefore the base elements Co and Ni are mixed in the alloys NC25, NC50 and NC75. Therefore, we propose a method to use mean values weighted according to the Co-content $f_B(Co)$ to determine the effective diffusion coefficients of these alloys. For $D_0$ we took harmonic mean values of the frequency factors of the solutes in Co and Ni, respectively, which was calculated as

$$\overline{D}_0^{i,Co/Ni} = \left( \left( \frac{1-f_B}{D_0^{i,Ni}} \right) + \left( \frac{f_B}{D_0^{i,Co}} \right) \right)^{-1}. \qquad \text{Equation A1}$$

For the activation energies of the alloying elements we used the weighted arithmetic mean of the $Q$ values of the individual solutes in Ni and Co, respectively:

$$\overline{Q}^{i,Co/Ni} = (1-f_B)Q^{i,Ni} + f_B Q^{i,Co} \qquad \text{Equation A2}$$

The determination of the activation energy of the base system is more complicated since one has to consider the activation energies for diffusion of Co in Co, Ni in Ni, Ni in Co and Co in Ni. Thus, we calculated a weighted mean of self-diffusion and the diffusion of the equivalent counterpart element:

$$\overline{Q}^{Co,Co/Ni} = (1-f_B)Q^{Co,Ni} + f_B Q^{Co,Co} \qquad \text{Equation A3}$$

$$\overline{Q}^{Ni,Co/Ni} = (1-f_B)Q^{Ni,Ni} + f_B Q^{Ni,Co} \qquad \text{Equation A4}$$

To get one value for $Q_{base}$ (see equation 5 in the manuscript) from $\bar{Q}^{Co,Co/Ni}$ and $\bar{Q}^{Ni,Co/Ni}$ we again calculated a weigthed arithmetic mean:

$$Q_{base} = (1-f_B)\bar{Q}^{Ni,Co/Ni} + f_B\bar{Q}^{Co,Co/Ni} \qquad \text{Equation A5}$$

The combination of equations A3, A4 and A5 results in:

$$Q_{base} = (1-f_B)^2 Q^{Ni,Ni} + (f_B - f_B^2)(Q^{Ni,Co} + Q^{Co,Ni}) + f_B^2 Q^{Co,Co} \qquad \text{Equation A6}$$

The calculated activation energies $Q_{base}$ for the NCX alloys, acquired by the procedure presented above, are presented in Figure A1. For NC0 ($f_B = 0$) and NC100 ($f_B = 1$), equation A6 gives $Q^{Ni,Ni}$ and $Q^{Co,Co}$, respectively. As described above, the values for the intermediate alloys are weighted according to self-diffusion and diffusion of the equivalent counterpart Co/Ni in a way that, for example, all four values $Q^{Co,Co}$, $Q^{Ni,Ni}$, $Q^{Co,Ni}$ and $Q^{Ni,Co}$ contribute equally when $f_B = 0.5$. Using $Q_{base}$ for the NCX alloys calculated in this way, we could obtain the effective diffusion coefficients of the matrix compositions as given in Figure 5d.

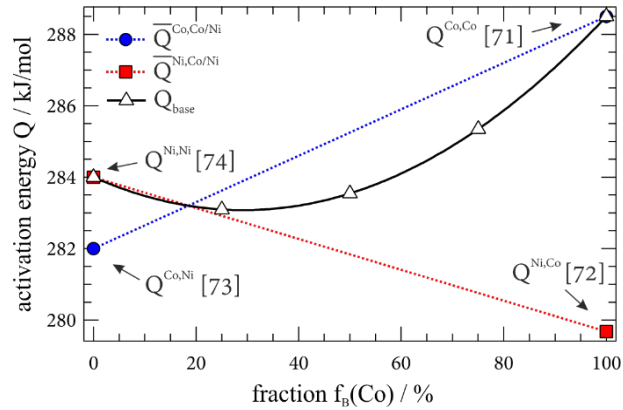

**Figure A1: Activation energies $Q_{base}$ for the NCX alloys according to the procedure proposed in the appendix. Activation energies from literature for Co in Co [79], Ni in Co [80], Co in Ni [81] and Ni in Ni [82] are shown as comparison.**

[14] A.B. Parsa, P. Wollgramm, H. Buck, A. Kostka, C. Somsen, A. Dlouhy, G. Eggeler, Ledges and grooves at γ/γ' interfaces of single crystal superalloys, Acta Mater. 90 (2015) 105–117. https://doi.org/10.1016/j.actamat.2015.02.005.

[15] M. Feller-Kniepmeier, T. Link, Dislocation structures in γ-γ' interfaces of the single-crystal superalloy SRR 99 after annealing and high temperature creep, Mater. Sci. Eng. A. 113 (1989) 191–195. https://doi.org/10.1016/0921-5093(89)90306-7.

[16] H. Mughrabi, Microstructural aspects of high temperature deformation of monocrystalline nickel base superalloys: some open problems, Mater. Sci. Technol. 25 (2009) 191–204. https://doi.org/10.1179/174328408X361436.

[17] C.M.F. Rae, R.C. Reed, Primary creep in single crystal superalloys: Origins, mechanisms and effects, Acta Mater. 55 (2007) 1067–1081. https://doi.org/10.1016/j.actamat.2006.09.026.

[18] T.M. Pollock, A.S. Argon, Creep resistance of CMSX-3 nickel base superalloy single crystals, Acta Metall. Mater. 40 (1992) 1–30. https://doi.org/10.1016/0956-7151(92)90195-K.

[19] G.R. Leverant, B.H. Kear, J.M. Oblak, Creep of precipitation-hardened nickel-base alloy single crystals at high temperatures, Metall. Trans. 4 (1973) 355–362.

[20] F. Pyczak, B. Devrient, H. Mughrabi, The effects of different alloying elements on the thermal expansion coefficients, lattice constants and misfit of nickel-based superalloys investigated by X-ray diffraction, Superalloys 2004. (2004) 827–836.

[21] A. Royer, P. Bastie, M. Veron, In situ determination of γ' phase volume fraction and of relations between lattice parameters and precipitate morphology in Ni-based single crystal superalloy, Acta Mater. 46 (1998) 5357–5368. https://doi.org/10.1016/S1359-6454(98)00206-7.

[22] C. Schulze, M. Feller-Kniepmeier, Transmisson electron microscopy of phase composition and lattice misfit in the Re-containing nickel-base superalloy CMSX-10, Mater. Sci. Eng. A. 281 (2000) 204–212. https://doi.org/10.1016/S0921-5093(99)00713-3.

[23] S. Neumeier, F. Pyczak, M. Göken, The temperature dependent lattice misfit of rhenium and ruthenium containing nickel-base superalloys – Experiment and modelling, Mater. Des. 198 (2021) 109362. https://doi.org/10.1016/j.matdes.2020.109362.

[24] C.H. Zenk, S. Neumeier, H.J. Stone, M. Göken, Mechanical properties and lattice misfit of γ/γ' strengthened Co-base superalloys in the Co–W–Al–Ti quaternary system, Intermetallics. 55 (2014) 28–39. https://doi.org/10.1016/j.intermet.2014.07.006.

[25] N. Volz, C.H. Zenk, R. Cherukuri, T. Kalfhaus, M. Weiser, S.K. Makineni, C. Betzing, M. Lenz, B. Gault, S.G. Fries, J. Schreuer, R. Vaßen, S. Virtanen, D. Raabe, E. Spiecker, S. Neumeier, M. Göken, Thermophysical and mechanical properties of advanced single crystalline Co-base superalloys, Metall. Mater. Trans. A. 49 (2018) 4099–4109. https://doi.org/10.1007/s11661-018-4705-1.

[26] U. Tetzlaff, H. Mughrabi, Enhancement of the high-temperature tensile creep strength of monocrystalline nickel-base superalloys by pre-rafting in compression, Pollock TM Kissinger RD Bowman RR Green KA McLean M. (2000). http://www.tms.org/superalloys/10.7449/2000/Superalloys_2000_273_282.pdf (accessed August 8, 2017).

[27] H. Mughrabi, W. Schneider, V. Sass, C. Lang, The effect of raft formation on the high-temperature creep deformation behaviour of the monocrystalline nickel-base superalloy CMSX-4, Strength Mater. ICSMA 10. (1994) 705–708.

# Supplementary material for:

# Creep properties and deformation mechanisms of single-crystalline γ′-strengthened superalloys in dependence of the Co/Ni ratio

N. Volz[a], C.H. Zenk[a], N. Karpstein[b], M. Lenz[b], E. Spiecker[b], M. Göken[a], S. Neumeier[a]

[a] Friedrich-Alexander-Universität Erlangen-Nürnberg, Department of Materials Science & Engineering, Institute I: General Materials Properties, 91058 Erlangen, Germany

[b] Friedrich-Alexander-Universität Erlangen-Nürnberg, Department of Materials Science & Engineering, Institute of Micro- and Nanostructure Research, and Center for Nanoanalysis and Electron Microscopy (CENEM), Friedrich Alexander University of Erlangen-Nuremberg, IZNF 91058 Erlangen, Germany

________________________________________________________________________

**Part A: Quantification of solid solution strengthening using different atomic radii for Al**

The manuscript describes and discusses the solid solution strengthening of the matrix compositions as determined by a model that was originally proposed by Labusch [1] and later modified by Gypen and Deruyettere [2,3] and Galindo-Nava et al. [4]. It was briefly mentioned in the manuscript that the atomic radii, which are used for the calculations, need to be chosen carefully. This became evident in the comparison of the results of this study and the findings of Galindo-Nava et al. [4]. Here, especially the atomic radii used for the calculations are different, especially the one of Al which was assumed to be 0.143 nm at room temperature (0.147 nm at 900 °C) compared to 0.124 nm in [4] (0.128 nm at 900 °C, assuming identical thermal behavior). We chose to use the metallic atomic radii, as they are also implemented in the used Thermo-Calc (TC) database equally for every alloying element, whereas the value used in [4] is closer to the covalent atomic radius of Al. To clarify especially the influence of the atomic radii of Al, where different values of the radius are used in literature, we also conducted the same calculations with $r_{Al}$=0.124 at room temperature, the value used by Galindo-Nava et al [4]. The results are shown in Figure S 1 in gray symbols and dashed lines in comparison with the values described and discussed in the manuscript (solid lines). Of course, the weighted solid solution strengthening is calculated to be lower since the atomic radii differences between Al and Ni (the reference element) is smaller in this case, however, the trend stays the same. Also with the atomic radius calculated to be 0.128 nm at 900 °C, the solid solution strengthening is decreasing on the Co-rich side, since the dominant factor is the γ and γ′ fraction of the alloy. Contrary, the isolated γ solid solution strength is calculated to decrease with increasing Co-content as well using $r_{Al}$=0.128 nm, whereas it was proposed to increase using $r_{Al}$=0.147 nm. This indicates that the atomic radii values, which are used in the model by Gypen and Deruyttere [2,3], have to be considered carefully. However, weakening the influence of Al using $r_{Al}$=0.128 nm is in agreement with experimental results on the lattice parameter change in Ni-Al binary systems that propose a significantly smaller impact of Al compared to W [5]. This would result in less lattice strains and therefore in lower solid solution strengthening. As a consequence, it might be necessary to adjust the models proposed by Fleischer [6], Gypen and Deruyttere [2,3] and Galindo-Nava et al. [4] in a way that not the pure element atomic radii are used. Instead the atomic volumes, which a solute exhibits in a binary system with the base element, as it could be calculated by TC for example, might be more suitable.

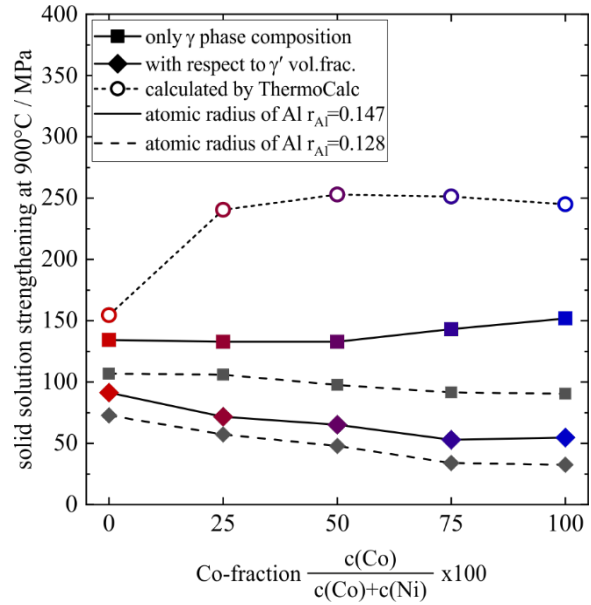

Figure S 1: Solid solution strengthening contribution of the γ phase at 900 °C calculated after [2–4] for the pure γ phase composition (squares, solid line), with respect to the γ′ volume fraction (diamonds, solid line). These calculations are repeated with a different atomic radius for Al of 0.128 nm as used in [4] (grey squares and diamonds, dashed line). Additionally, Thermo-Calc was used to calculate the solid solution strengthening using different model as proposed by Walbrühl et al. [7] (open circles).

**Part B: Quantification of the solid solution strengthening using Thermo-Calc**

The solid solution strengthening of the matrix phase was also calculated directly using the TCNI10 database and the corresponding property model implemented in Thermo-Calc 2021a. Internally, this model is based on Walbrühl et al. [7], who assume the same concentration dependence as Labusch [1], but fit their model directly to experimental hardness data. As before, the matrix composition as determined from APT reconstructions was used. σss was evaluated at the creep test temperature of 900 °C. The results are also shown in Figure S 1 (open circles).

Similar to the results described in the manuscript and above, an increasing solid solution strengthening contribution with increasing Co-content is found. However, the absolute values differ significantly. Especially the increase from the pure Ni-alloy NC0 to the Co-containing alloy NC25 is much more pronounced, which is consistent with the increasing creep resistance between these two alloy compositions. Furthermore, the Thermo-Calc property model reveals a maximum in the strengthening effect for NC50 while our calculations applying the model by Gypen and Deruyttere [2,3] estimate a steady increase with a maximum for NC100. The model used by TC was derived by Walbrühl et al. [7] and is also based on a model originally proposed by Labusch [1]. However, a non-linear composition dependence of the strengthening parameter is applied in their model and they fit their model directly to experimental hardness data, which is different to the model by Gypen and Deruyttere [2,3]. We assume that these differences in the models cause the changing and non-conform calculations.